\documentclass{llncs}
\usepackage{llncsdoc}
\usepackage{graphicx}
\begin{document}

\author{R.~Farinelli$^a$$^b$ *,M.~Alexeev$^c$, A.~Amoroso$^c$, F.~Bianchi$^c$, M.~Bertani$^d$, D.~Bettoni$^a$, N.~Canale$^a$$^b$, A.~Calcaterra$^d$, V.~Carassiti$^a$, S.~Cerioni$^d$, J.~Chai$^c$,  S.~Chiozzi$^a$, G.~Cibinetto$^a$, A.~Cotta Ramusino$^a$, F.~Cossio$^c$, F.~De~Mori$^c$, M.~Destefanis$^c$, T.~Edisher$^d$, F.~Evangelisti$^a$, L.~Fava$^c$, G.~Felici$^d$, E.~Fioravanti$^a$, I.~Garzia$^a$$^b$, M.~Gatta$^d$, M.~Greco$^c$, D.~Jing$^d$, L.~Lavezzi$^c$$^e$, C.~Leng$^c$, H.~Li$^c$, M.~Maggiora$^c$, R.~Malaguti$^a$, S.~Marcello$^c$, M.~Melchiorri$^a$, G.~Mezzadri$^a$$^b$, G.~Morello$^d$,S.~Pacetti$^f$, P.~Patteri$^d$, J.~Pellegrino$^c$, A.~Rivetti$^c$, M.~D.~Rolo$^c$, M.~Savrie'$^a$$^b$, M.~Scodeggio$^a$$^b$, E.~Soldani$^d$, S.~Sosio$^c$, S.~Spataro$^c$, L.~Yang$^c$.
\\ ~
\institute{$^a$ INFN - Sezione di Ferrara, $^b$ University of Ferrara,$^c$ INFN - Sezione di Torino,\\ $^d$ INFN - Sezione di Frascati,  Physics dept., $^e$ IHEP, Beijing, \\$^f$ University of Perugia.}
\\E-mail address: rfarinelli@fe.infn.it (R.~Farinelli), *Corresponding author.}%

\title{A Cylindrical GEM Inner Tracker for the BESIII experiment at IHEP}{A Cylindrical GEM Inner Tracker for the BESIII experiment at IHEP}
\maketitle
\begin{abstract}
The Beijing Electron Spectrometer III (BESIII)\cite{bes3} is a multi-purpose detector that collects data provided by the collision in the Beijing Electron Positron Collider II (BEPCII), hosted at the Institute of High Energy Physics of Beijing. Since the beginning of its operation, BESIII has collected the world largest sample of J/$\psi$ and $\psi$(2s). Due to the increase of the luminosity up to its nominal value of $10^{33}$ cm$^{-2}$s$^{-1}$ and aging effect, the MDC decreases its efficiency in the first layers up to 35$\%$ with respect to the value in 2014. Since BESIII has to take data up to 2022 with the chance to continue up to 2027, the Italian collaboration proposed to replace the inner part of the MDC with three independent layers of Cylindrical triple-GEM (CGEM).

The CGEM-IT project will deploy several new features and innovation with respect the other current GEM based detector: the $\mu$TPC and analog readout, with time and charge measurements will allow to reach the 130 $\mu$m spatial resolution in 1 T magnetic field requested by the BESIII collaboration.
In this proceeding, an update of the status of the project will be presented, with a particular focus on the results with planar and cylindrical prototypes with test beams data. These results are beyond the state of the art for GEM technology in magnetic field.
\end{abstract}
%
%
%
\section{Introduction}
GEM technology\cite{sauli} exploits 50/70 $\mu$m holes on a 50 $\mu$m kapton foil with 3 $\mu$m copper on the faces. An high voltage of hundreds of Volts creates an intense field that can creates an electron avalanche if an electron passes through the hole. The use of three GEM foils allows to achieve a gain of 10$^3$-10$^4$ with a low discharge probability and this assures the electrical stability of the detector\cite{disch}. A cathode and a stripped anode ultimate the design. The CGEM detector introduces the challenge to shape a large area triple-GEM. The performance of this detector has to be better than the preexistent one. Moreover it will introduce a series of benefits such as the improvement of the spatial resolution of the coordinate along the beam direction, i.e. an improvement of a factor 2-3 of the vertex resolution for $K_s^0$ and $\Lambda$ is expected.
%
\section{Planar GEM Results}
\begin{figure}
\includegraphics[width=0.5\textwidth]{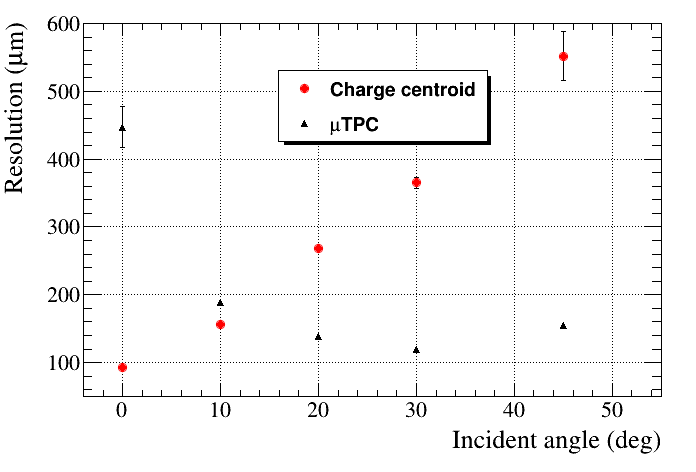}
\includegraphics[width=0.5\textwidth]{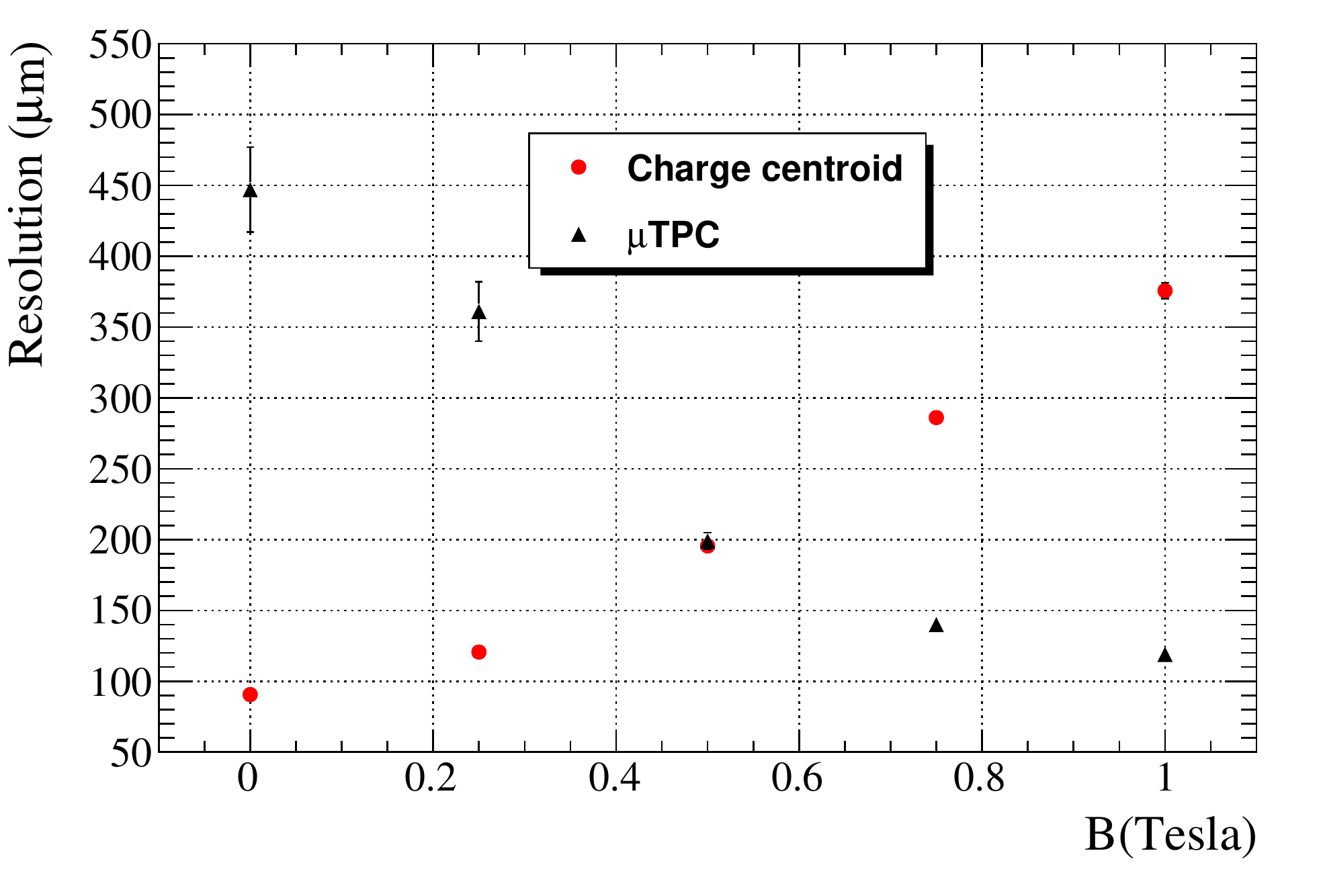}
\vspace{0.2cm}
\caption{CC and $\mu$TPC spatial resolution as a function of the incident angle (left) and the magnetic field (right). 0$^{o}$ corresponds to orthogonal tracks.}
\label{scan}
\end{figure}

A characterization of the triple-GEM with 10$\times$10 cm$^2$ planar detectors has been performed with a test beam campaign with muon and pion beams in the H4 line at SPS(CERN)\cite{tb}. The detector performance strictly depends on the geometrical arrangement, gas mixture and the applied fields. Studies with Argon:CO$_2$ (70:30) and Argon:iC$_4$H$_{10}$ (90:10) have been performed. The second one has been chosen to obtain an higher electrical stability, higher number of primary electron and a suitable diffusion properties for the optimization of the reconstruction algorithms: the charge centroid and the micro-Time Projection Chamber ($\mu$TPC). The gap between the cathode and the first GEM is a parameter that influences the charge collection: a 5 mm conversion gap  has been chosen to reach good performance with angled tracks. The gain is set by the voltage applied between the GEM faces and at about a gain of 6000 the detector reaches the efficiency plateau of 97$\%$ on the two views. As the gain increases, the signal charge and the number of fired strips increase. To reconstruct the position, the charge and the time measured by each strip are used. The charge centroid calculates the weighted average position of the strips with their charge. Its best performance is achieved when the cluster size is greater than two, without magnetic field and for orthogonal incident tracks. In this case the resolution obtained with this method is stable as the cluster size increases since the collected charge distribution has a Gaussian shape. The time based algorithm, the $\mu$TPC, improves the detector resolution in presence of magnetic field and non-perpendicular tracks. Using the drift velocity from Garfield simulation it is possible to assign to each fired strip a bi-dimensional point. 
Once the working point is reached and the detector is efficient, measurements of its performance as a function of the incident angle and the magnetic field are proposed to clarify the behavior of the detector. It is important to separate the signal formation in two steps: the former where the primary electrons are generated and reach the first GEM foil, the latter where each electron is amplified three times and a Gaussian charge distribution is collected on the anode per each primary electron. the results as a function of the incident angle in absence of magnetic field show the CC provides a resolution better than 100 $\mu$m at 0$^{o}$ (see Fig. \ref{scan} left). As the incident angle increases the charge distribution charges from a Gaussian shape to a box shape at large angle and this degrades the CC resolution. Conversely, as the incident angle increases, the $\mu$TPC improves because the number of fired strips increases. It is important to evidence that the two reconstruction algorithms are anti-correlated and one gives its best performance when the other is inefficient.
\\
A similar behavior of the performances is measured as a function of the magnetic field (Fig. \ref{scan} right). Here the Lorentz force drifts the electrons and enlarges the charge distribution at the anode. The Lorentz angle is the angle within the track path and the avalanche direction. Without magnetic field the CC has a resolution better than 100 $\mu$m but as the magnetic field increases, the Lorentz angle increased too and the efficient algorithm becames the $\mu$TPC. 
The final measurement proposed to characterize a $\mu$TPC it is the performance measurement as a function of the incident angle at 1 T (Fig. \ref{otro} left). Here the combination of the track angle and the Lorentz angle creates a pattern at the anode that can be described with two configurations: focusing and defocussing. If there is focusing effect, the incident angle coincides with the Lorentz angle, the avalanche distribution is Gaussian and the CC gives it best performance. The Lorentz angle of these setting is 26$^o$.  As the incident angle moves away from the Lorentz angle the CC degrades and the $\mu$TPC achieves its best performance. A combination of the two methods should keep the resolution stable around 130 $\mu$m in the full range of the incident angles.
\section{CGEM Characterization and Results}
\begin{figure}
\includegraphics[width=0.5\textwidth]{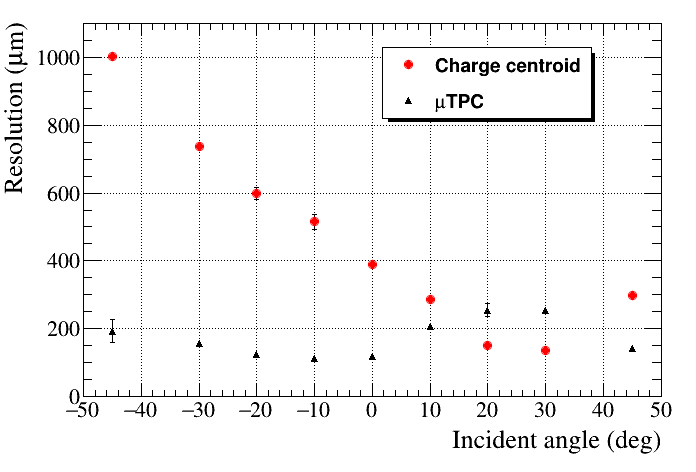}
\includegraphics[width=0.5\textwidth]{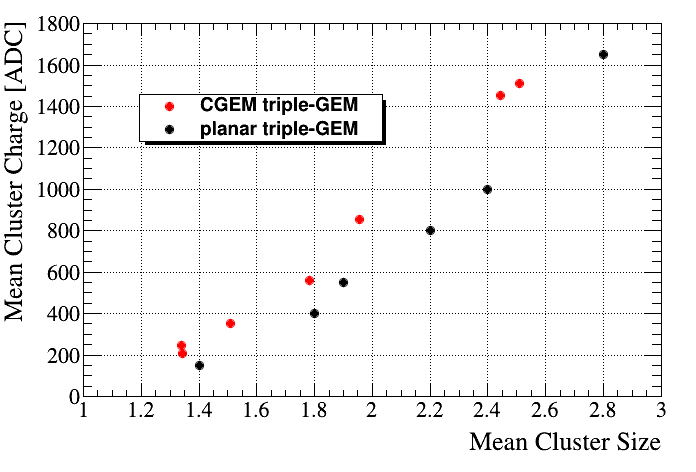}
\vspace{0.2cm}
\caption{Left) CC and $\mu$TPC in 1 T magnetic field as a function of the indicent angle. 0$^{o}$ corresponds to orthogonal tracks. Right) Mean charge collected by a planar GEM and a CGEM as a function of the mean cluster size. The data have been collected for different gain values.}
\label{otro}
\end{figure}

A large area triple-GEM with cylindrical shape has been built in the Ferrara and Frascati INFN workshops and in order to validate the construction it has been tested in a test beam at CERN. The built CGEM has a radius of 20 cm and a length of 80 cm and the first two challenge to achieve its operability are the gas sealing and the electrical stability. A study of the collected charge as a function of the mean cluster size has been performed for different values of gain. Using orthogonal tracks and no magnetic field the CGEM shows the same linearity of the planar GEM (Fig. \ref{otro} right). This is a clear indication that the behavior of the electron avalanche inside the detectors is the same and the ratio between the charge and the number of the fired strip in the CGEM is compatible with the planar one. This grants the applicability of the reconstruction algorithms developed for the planar chambers also to the cylindrical GEM. A spatial resolution of 110 $\mu$m has been measured with the CC. The $\mu$TPC algorithm needs angled tracks or magnetic field to be tested. This study will be performed in the next test beam.


\section{References}
\begin{thebibliography}{[MT1]}
%
\bibitem{bes3}
BESIII Collaboration, \emph{The construction of the BESIII experiment, Nucl. Instr. Meth. A 598 (2009)} 7-11000000
\bibitem{sauli}
F.~Sauli, \emph{Nucl. Instrum. Methods A}, 386:531, 1997
\bibitem{disch}
S. Bachmann et al, \emph{Discharge studies and prevention in the gas electron multiplier (GEM), Nucl. Instr.} and Meth. A 479 (2002) 294
\bibitem{tb}
cern.ch/RD51-Public/
\end{thebibliography}
\end{document}